Review

# Routes for breaching and protecting genetic privacy

Yaniv Erlich[1,*] and Arvind Narayanan[2]

[1] Whitehead Institute for Biomedical Research, Nine Cambridge Center, Cambridge, MA USA 02142
[2] Department of Computer Science, Princeton University, 35 Olden Street, Princeton, NJ USA 08540
[*] Correspondence to Y.E (yaniv@wi.mit.edu)

## Abstract

We are entering the era of ubiquitous genetic information for research, clinical care, and personal curiosity. Sharing these datasets is vital for rapid progress in understanding the genetic basis of human diseases. However, one growing concern is the ability to protect the genetic privacy of the data originators. Here, we technically map threats to genetic privacy and discuss potential mitigation strategies for privacy-preserving dissemination of genetic data.

## About the Authors

Yaniv Erlich is a Fellow at the Whitehead Institute for Biomedical Research. Erlich received his Ph.D. from Cold Spring Harbor Laboratory in 2010 and B.Sc. from Tel-Aviv University in 2006. Prior to that, Erlich worked in computer security and was responsible for conducting penetration tests on financial institutes and commercial companies. Dr. Erlich's research involves developing new algorithms for computational human genetics.

Arvind Narayanan is an Assistant Professor in the Department of Computer Science and the Center for Information Technology and Policy at Princeton. He studies information privacy and security. His research has shown that data anonymization is broken in fundamental ways, for which he jointly received the 2008 Privacy Enhancing Technologies Award. His current research interests include building a platform for privacy-preserving data sharing.

## Summary

- Broad data dissemination is essential for advancements in genetics, but also brings to light concerns regarding privacy.

- Privacy breaching techniques work by cross-referencing two or more pieces of information to gain new, potentially undesirable knowledge on individuals or their families.

- Broadly speaking, the main routes to breach privacy are identity tracing, attribute disclosure, and completion of sensitive DNA information.

- Identity tracing exploits quasi-identifiers in the DNA data or metadata to uncover the identity of an unknown genetic dataset.

- Attribute disclosure techniques work on known DNA datasets. They use the DNA information to link the identity of a person with a sensitive phenotype.

- Completion techniques also work on known DNA data. They try to uncover sensitive genomic areas that were masked to protect the participant.

- In the last few years, we have witnessed a rapid growth in the range of techniques and tools to conduct these privacy-breaching attacks. Currently, most of the techniques are beyond the reach of the general public, but can be executed by trained persons with varying degrees of effort.

- There is considerable debate regarding risk management. One camp supports a pragmatic, ad-hoc approach of privacy by obscurity and the other supports a systematic, mathematically-backed approach of privacy by design.

- Privacy by design algorithms include access control, differential privacy, and cryptographic techniques. So far, data custodians of genetic databases mainly adopted access control as a mitigation strategy.

- New developments in cryptographic techniques may usher in an additional arsenal of security by design techniques.

# INTRODUCTION

We produce genetic information for research, clinical care, and genealogy at exponential rates. Sequencing studies with thousands of individuals have become a reality[1,2] and new projects aim to sequence hundreds of thousands to millions of individuals[3]. Some geneticists envision whole genome sequencing of every person as part of routine health care[4,5].

Sharing genetic findings is vital for accelerating the pace of biomedical discoveries and fully realizing the promises of the genetic revolution[6]. Recent studies suggest that robust predictions of genetic predispositions to complex traits from genetic data will require the analysis of millions of samples[7,8]. Clearly, collecting cohorts at such scales are typically beyond the reach of individual investigators and cannot be achieved without combining different sources. In addition, broad dissemination of genetic data promotes serendipitous discoveries through secondary analysis, which is necessary to maximize its utility for patients and the general public[9].

One of the key issues of broad dissemination is an adequate balance of data privacy[10]. Prospective participants of scientific studies have ranked privacy of sensitive information as one their top concerns and a major determinant if to participate in the study[11-13]. Protecting personally identifiable information is also a demand of an array of regulatory statutes in United States and the European Union[14]. Data de-identifying, the removing of the person identifier, has been suggested as a potential path to reconcile data sharing and privacy demands[15]. But is this technically feasible for genetic data?

This review characterizes privacy breaching techniques of genetic information and maps potential counter-measures. We first categorize privacy-breaching strategies, discuss their underlying technical concepts, and evaluate their performance and limitations. Then, we present privacy-preserving technologies, group them according to their methodological approaches, and discuss their relevance to genetic information. As a general theme, we focus only on breaching techniques that involve data mining and fusing distinct resources to gain private information relevant to DNA data. Data custodians should be aware that security threats can be much broader. They can include cracking weak database passwords, classical computer hacking techniques of the server that holds the data, stealing of storage devices due to poor physical security, and intentional misconduct of data custodians[16-18]. We do not include these threats since they are not unique to genetic information and have been extensively studied by the computer security field[19]. In addition, this review does not cover the potential implications of loss of privacy, which heavily depend on cultural, legal, and socio-economical context and were covered in part by the broad privacy literature[20,21].

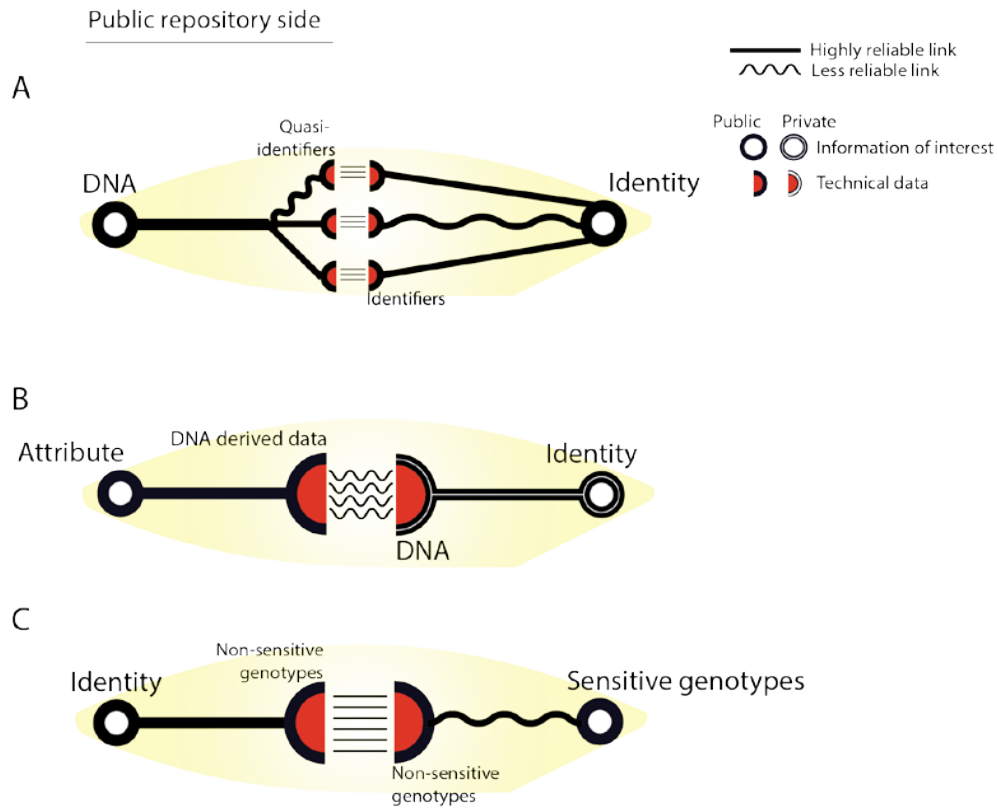

**Figure 1 | Main routes for breaching genetic privacy.** On the left, schematic representation of various datasets that are hosted by scientific repositories. On the right, the conjugate datasets that match to each representation route. a | Identity tracing aims to create a statistical bridge between DNA data and identities using quasi-identifiers. b | Attribute Disclosure Attacks via DNA (ADAD) match between identities and sensitive phenotypes using DNA derived data as a linker. c | Completion techniques enable revealing sensitive genotypes of a specific identity using reference panels.

## PRIVACY BREACHING OF GENETIC DATA

Genetic privacy breaching techniques fall into three categories: Identity Tracing, Attribute Disclosure Attacks via DNA (ADAD), and Completion Techniques (**Figure 1**). The shared concept of these techniques is gaining a new piece of private – potentially sensitive – information about the target or his family by exploiting DNA data. The three categories are distinct in the type of sensitive information that they reveal. The aim of identity tracing is to link between an unknown genome and the concealed identity of the data originator. In ADAD, the adversary already has access to the identified DNA sample of the target and to a database that links DNA-derived data to sensitive attributes without explicit identifiers, for example a public database of the genetic study of drug abuse. The ADAD techniques match the DNA data and associate the identity of the target with the sensitive attribute. In completion techniques, the adversary also knows the identity of a genomic dataset but has access only to a sanitized version without sensitive loci. The aim here is to expose the sensitive loci that are not part of the original data. **Table 1** summarizes all privacy breaching techniques that are presented in this section.

Table 1 | **Categorization of techniques for breaching genetic privacy**

| Technique | Maturation Level | Technical complexity | Auxiliary information |
|---|---|---|---|
| **Identity Tracing** | | | |
| Surname Inference | ★★★★ | ••• | Intermediate-Good |
| DNA Phenotyping | ★★ | •• | Poor |
| Demographic identifiers | ★★★★ | • | Good |
| Pedigree structure | ★★★ | •• | Poor |
| Side channel leakage | ★★★★ | ••• | Varies |
| **Attribute Disclosure Attacks via DNA** | | | |
| N=1 | ★★★★ | ••• | Good |
| Genotype frequencies | ★★★ | ••• | Good |
| Linkage disequilibrium | ★★ | •••• | Intermediate |
| Effect sizes | ★★ | ••• | Good |
| Trait inference | ★ | •• | Good |
| Gene expression data | ★★★ | •••• | Poor |
| **Completion Attacks** | | | |
| Imputation of a masked marker | ★★★★ | •• | Good |
| Genealogical imputation | ★★★★ | •••• | Poor |

Maturation level: *Working principles established with simulated data. **Small scale proof of concept with real data in a controlled environment (typically only one dataset). ***Large scale experiments in controlled environments with real data (typically more than one dataset). ****Breach of privacy was reported in a real scenario.

Technical complexity: • no knowledge in genetics or special tools is required. ••Require genetic knowledge; computation can reasonably be done on a regular computer. Existing tools are available •••Require genetic knowledge, intermediate scale processing of data and/or molecular techniques. ••••Require genetic knowledge; large scale processing of data is a prerequisite; may also require molecular techniques.

Auxiliary information: this column refers to the level of existing reference databases for the US population in public resources. For identity tracing, it refers to the availability of organized lists that link identities and extract pieces of information. For ADAD and completion techniques, it refers to the existence of supporting reference datasets that are necessary to complete the attack. Poor – supporting data is highly fragmented and not amenable to searching. Intermediate – supporting data is harmonized and searchable but requires some pre-processing. Great – supporting data is ready to use using existing tools or minimal pre-processing.

## IDENTITY TRACING ATTACKS

The goal of identity tracing attacks is to uniquely identify the data originator from the population despite the absence of explicit identifiers such as the name and exact address in the published dataset. The idea is to accumulate quasi-identifiers -- residual pieces of information that are embedded in the dataset -- and to gradually narrow down the possible individuals that match the combination of these quasi-

## Box 1 | Entropy and the contribution of quasi-identifiers

Entropy is the measure of the degree of uncertainty in the outcome of a random variable. One bit of entropy is equivalent to the uncertainty of tossing a fair coin, two bits of entropy are equivalent to two independent tosses of a fair coin and so on. Zero bits is the lowest entropy level and implies that there is no uncertainty. The reciprocal measure of entropy is information content, which quantifies the contribution of a new piece of data to the entropy level.

In the study of data anonymization, entropy expresses the expected level of uncertainty of a potential match between an individual and the dataset. Consider an anonymous individual's record in a study that randomly samples subjects from the US population. A priori, the adversary has 310 million equiprobable options of a match, which translates to 28.2bits of entropy. He can then gain 1 bit of information by inferring the individual's sex and reduce the entropy to 27.2. Complete identification is achieved by reducing the entropy to zero. The **table** below lists possible quasi-identifiers and their maximal information content expectation for the US population.

Several factors reduce the expected information content of quasi-identifiers from the maximal level. One possibility is that two quasi-identifiers are correlated. For example, after inference of the zip code, obtaining information on state adds no new information. A second possibility is an inaccurate inference of the quasi-identifier. Information theory dictates a rapid decline of information content with deviations of the inferred quasi-identifier from the truth (see **figure**). Another possibility is low-searchability of the quasi-identifier. For example, in the case that the adversary can only access a height registry of 100 random US individuals, even with perfect knowledge of height, he will recover close to zero bits of information. Relying on low searchability to prevent privacy breach might not be sustainable. The social media ecosystem can quickly establish new types of highly searchable registries without data custodian control, rendering the data vulnerable.

| Quasi-identifier | Expected information content (bits) |
|---|---|
| Sex[1] | 1.0 |
| Ethnic group[1,2] | 1.4 |
| Eye color[3] | 1.4 |
| Blood group (ABO/Rh)[4] | 2.2 |
| State[1] | 5.0 |
| Height[5] | 5.0 |
| Year of birth[1] | 6.3 |
| Day and month of birth[6] | 8.5 |
| Surname[1] | 12.9 |
| Zip code[7] | 13.8 |
| Match of autosomal DNA[8] | 28.2 |

Table: Information content of quasi-identifiers for the general US population

[1] Based on the US Census data.
[2] Based on self-classification field in the US census: African American, Asian American, European American, Native American, Other race, and two or more races.
[3] Perfect inferences of three eye colors groups (blue, brown, intermediate). Data from www.statisticbrain.com/eye-color-distribution-percentages/
[4] Data is based on Stanford School of Medicine Blood Center (bloodcenter.stanford.edu/about_blood/blood_types.html)
[5] Assuming accurate measure in 1cm resolution and 8cm standard deviation in the population.
[6] Data is based on 400,000 births (www.panix.com/~murphy/bday.html)
[7] Data is from zipatals.com
[8] Excluding monozygotic twins.

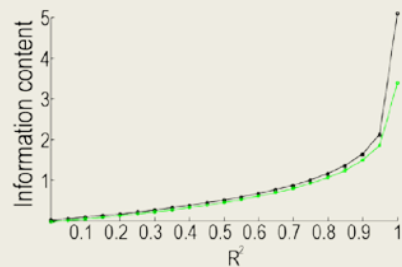

Figure Box 1 | **The information content of noisy quasi-identifiers.** The two quasi-identifiers distribute normally, black: 100 levels, green: 32 levels. $R^2$ denotes the coefficient of determination between the true values of the quasi-identifier and the inferred values. In low $R^2$ values, the information content is very small regardless of the number of levels.

identifiers to the point that the data originator is the only match. The success of the attack depends on the information content that the adversary can obtain from these quasi-identifiers relative to size of the base population (**Box 1**).

## IDENTITY DISCLOSURE BY META-DATA

Genetic datasets are typically published with additional metadata, such as basic demographic details, inclusion/exclusion criteria, pedigree structure, and health conditions that are critical to understand the study and for secondary analysis. Unrestricted demographic information conveys substantial power for identity tracing. It has been estimated that the combination of date of birth, sex, and 5 digit zip code uniquely identifies more than 60% of US individuals[22,23]. In addition, there are extensive public resources with broad population coverage and search interfaces that link demographic quasi-identifiers and individuals, including voter registries, public record search engines such as People- Finders.com, data brokers, and social media. In one of the pioneering studies of identity tracing using metadata, Sweeny reported the successful tracing of the medical record of the Governor of Massachusetts using demographic identifiers[24]. At that time, the Massachusetts Group Insurance Commission released hospital discharge information with five digit zip codes, sex, and date of birth. By searching the voter registry, Sweeny was able to uniquely match the hospital discharge of the Governor. A more recent study reported the identification of 30% of Personal Genome Project (PGP) participants by demographic profiling that included zip code and exact birthdates that are found in PGP profiles[25].

Since the inception of the HIPAA Rule in 2003, demographic identifiers are the subjects of tight regulation in the US health care system[26]. The [SAFE HARBOR](#) provision requires that the maximal resolution of any date field, including birth and hospital admissions, will be in years. In addition, the maximal resolution of a geographical subdivision is the first three digits of a zip code (as long as there are more than 20,000 living in the regions that correspond to the three digit zip codes). Statistical analyses of the census data have found that the Safe Harbor provision provides reasonable immunity against identity tracing assuming that the adversary has access only to demographic identifiers. The combination of sex, age, ethnic group, and state is unique to less than 0.25% of the populations across all states[27]. An empirical study evaluated the re-identification of 15,000 records of Hispanic patients in the Chicago area that included year of birth, 3-digit zip code, and marital status (married/unmarried) by comparison to voting registry data[28]. The authors reported the correct identification of 2 out of the 15,000 records and estimated that less of 0.22% the population is exposed with this set of quasi-identifiers. These studies show that with access to only HIPAA redacted demographic quasi-identifiers, identity tracing is extremely hard.

Pedigree structures are another piece of metadata that are included in many genetic studies. These structures contain rich information, especially when large kinships are available[29]. The number of offspring, their birth order, and other familial events such as remarriage, create unique combinations of quasi-identifiers that quickly narrow down the search space. A systematic study analyzed the distribution of 2,500 two-generation family pedigrees that were sampled from obituaries of a town of 60,000 individuals[30]. The pedigrees were unsorted, meaning that only the number of male and female individuals in each generation was available. Despite this limited information, about 30% of the pedigree structures were unique, demonstrating the large information content that can be obtained from such data. Another feature of pedigrees for identity tracing is the combination of quasi-identifiers across records. For example, it is quite rare that a surname alone can identity an individual. However, the surname combination of a couple prior to their marriage is an extremely strong identifier. In addition, once a single individual in a pedigree is identified, it is easy to link the identities of the other relatives and their genetic datasets. The main limitation of identity tracing using pedigree structures is their low searchability. Perhaps one notable exception is Israel, where the entire population registry was leaked to the web in 2006 and allows the construction of multi-generation family trees of all Israeli citizens[31]. But in general due to their low searchability, the value of family trees for re-identification is mostly limited to manual verification of the potential identity of the target rather than a starting point of the process.

## IDENTITY TRACING BY GENEALOGICAL TRIANGULATION

Genetic genealogy attracts millions of individuals interested in their ancestry and discovering distant relatives. To that end, the community has developed impressive online platforms to search for genetic matches and connect individuals. These online resources can be exploited to triangulate the identity of an unknown genome.

One potential route of identity tracing is surname inference from Y-chromosome data[32,33]. In most societies, surnames are passed from father to son, creating a transient correlation with specific Y chromosome HAPLOTYPES[34,35]. The adversary can take advantage of the Y chromosome-surname correlation and compare the Y haplotype of the unknown genome to haplotype records in recreational genetic genealogy databases. A close match with a relatively short time to the most common recent ancestor (MRCA) would signal that the unknown genome likely has the same surname as the record in the database.

The power of surname inference stems from exploiting information from distant patrilineal relatives of the unknown genome. The association between surnames and Y-chromosomes usually spans dozens of generations, implying that every record in a genealogical database is capable of revealing the surnames of hundreds to thousands of males. A recent empirical study estimated that 10-14% of US Caucasian males from the middle and upper classes are subject to surname inference based on scanning the two largest Y-chromosome genealogical websites with a built-in search engine[33].

An inferred surname has tremendous power for identity tracing. Individual surnames are relatively rare in the population and in most cases a single surname is shared by less than 40,000 US males[33], which is equivalent to 12 bits of information. In terms of identification, successful surname recovery is very close to determining an individual's zip code. Another feature of surname inference is that surnames are highly searchable. From public record search engines to social networks, numerous online resources offer surname query interfaces, simplifying the adversary's efforts to complete the triangulation.

Surname inference has been utilized to breach genetic privacy in the past[36-39]. Several sperm donor conceived individuals and adoptees successfully used this technique on their own DNA to reveal the surnames of their ancestors, which eventually lead to the exposure of their biological families. This technique could also be applied to whole genome sequencing datasets. A recent study reported five successful surname inferences from Illumina datasets of three large families that were part of the 1000 Genomes project, which eventually exposed the identity of close to fifty research participants[33].

The main limitation of surname inference is that haplotype matching relies on comparing Y chromosome Short Tandem Repeats (Y-STRs). Currently, most sequencing studies do not routinely report these markers and the adversary would have to process large-scale raw sequencing files with a specialized tool, which is both time and resource consuming and requires bioinformatics experience [40]. Another complication is false identification of surnames and inference of surnames with spelling variants compared to the original surname. Eliminating incorrect surname hits necessitates access to additional quasi-identifiers such as pedigree structure and typically requires a few hours of manual work. Finally, the performance of surname inference varies between different socio-ethnic groups based on non-paternity rates, sociological norms of surname inheritance, and access of the group to recreational genealogy.

An open research question is the utility of non Y chromosome markers for genealogical triangulation. Websites such as Mitosearch.org and GedMatch.com run open searchable databases for matching mitochondrial and autosomal genotypes, respectively. Our expectation is that mitochondrial data will not be very informative for tracing identities. The resolution of mitochondrial searches is much lower due to its smaller size and the absence of highly polymorphic markers like Y-STRs, meaning that a large number of individuals would share the same mitochondrial haplotypes. In addition, most human societies do not exercise maternally inherited identifiers, reducing the utility of such searches. Autosomal searches on the other hand might be quite powerful. Genetic genealogy companies have started to market services for dense genome-wide arrays that enable relatively sufficient accuracy to identify distant relatives on the order of 3rd to 4th cousins[41]. These hits would reduce the search space to no more than a few thousand individuals[42]. The main challenge of this approach would be translating the genealogical match to a list of potential people. But with the growing interest in genealogy, this technique might be easier in the future and should be taken into consideration.

## IDENTITY TRACING BY PHENOTYPIC PREDICTION

Several reports on genetic privacy envisioned that phenotypic predictions from genetic data could serve as quasi-identifiers for identity tracing[43,44]. Twin studies have estimated high heritabilities for various visible traits such as height[45] and facial morphology[46]. In addition, recent studies showed that age prediction is possible from DNA specimens derived from blood samples[47,48]. But the applicability of these DNA-derived quasi-identifiers for identity tracing has yet to be demonstrated.

The major limitation of phenotypic prediction is the fast decay of the identification power with small inference errors (**Box 1**). Current genetic knowledge explains only a small extent of the phenotypic variability of most visible traits, such as height[49], BMI[50], and face morphology[51], significantly limiting their utility for identification. For example, perfect knowledge about height at one-centimeter resolution conveys 5 bits of information. However, with current genetic knowledge that explains 10% of height variability[49], the adversary learns only 0.15 bits of information. Predictions of most of the other visible traits are even worse, implying that their utility as quasi-identifiers would be quite low. The exceptions in visible traits are eye color[52] and age prediction[47]. Recent studies showed a prediction accuracy of 75%-90% of the phenotypic variability of these traits. But even these successes translate to approximately 3-4 bits of information. Another challenge for phenotypic prediction is the low searchability of most of these traits. There are no population-based registries of height, eye color, or face morphology and the adversary would have to invest substantial efforts to compile such a registry. However, with the advent of new types of social media, this barrier might be less significant in the future.

## IDENTITY TRACING BY SIDE-CHANNEL LEAKS

Side channel attacks exploit quasi-identifiers that are unintentionally encoded in the database building blocks and structure rather than the actual data that is meant to be public. A good example for such leaks is the exposure of the full names of PGP participants from 23andMe filenames[25]. The PGP allowed participants to upload 23andMe genotyping files to their public profile webpages. The default convention of these 23andMe files includes the first and last name of the user. As part of the upload process, the PGP website automatically compressed the file, named it with the PGP identifier of the user, and presented a link that showed the new file name that does not include the first and last names. However, after downloading and decompressing the 23andMe file, the original filename appeared. Since most of the users did not change the default naming convention, it was possible to trace the identity of a large number of PGP profiles. Based on this experience, the PGP now forces the participant to rename files before uploading and warns them that the file may contain hidden information that can expose their identities.

Rich data files embed multiple layers of hidden information that provide ample opportunities for leakage of quasi-identifiers. Photo files typically embed Exchangeable Image File Format (EXIF) fields that can include GPS data about the location of the photo or the serial number of the camera[53]. This information could convey potential leads even if the photo itself does not disclose any sensitive information. Microsoft Office products typically embed the author name and contain previous revisions of the document that show deleted text[54]. In general, flat text files are the most immune format to these types of leaks of unintentional content.

The mechanism to generate database accession numbers can also leak personal information. Ideally, these numbers should be completely random but experience has highlighted that sometimes these numbers unintentionally reveal residual information due to non-random assignments. For example, in several top medial data mining contests, the accession numbers unintentionally revealed the disease status of the patient, which was the aim of the contest[55]. Another example is the non-random assignment of Social Security Numbers (SSN) in the US. Pattern analysis of a large amount of public data revealed temporal and spatial commonalities in the assignment system that allowed predictions of the SSN from quasi-identifiers[56]. Some suggested the assignment of accession numbers by applying CRYPTOGRAPHIC HASHING to the participant identifiers such as name or social security number[57]. However, this technique is extremely vulnerable to DICTIONARY ATTACKS due to the relatively low search space of the input. In general, it is advisable to add some sort of randomization to procedures that generate accession numbers in order to prevent misuse.

## ATTRIBUTE DISCLOSURE ATTACKS VIA DNA (ADAD)

In ADAD, the adversary creates a statistical bridge that uses DNA data to link sensitive attributes with the identity of a person. The first piece of information is a DNA sample from an identified target. This can be achieved by successful completion of an identity tracing attack, exploiting identified DNA data in projects such as OpenSNP, gaining internal access to restricted databases, or simply by obtaining a DNA sample directly from the target. The second piece of information is DNA derived data that is associated with sensitive information, such as disease, personality traits, or socio-economic status, which does not otherwise contain explicit identifiers. The main difference between the ADAD attacks is the type of DNA derived data that is associated with the sensitive attribute.

### ADAD: THE N=1 SCENARIO

The simplest scenario of ADAD is when the sensitive attribute is associated with the genotype data of the individual. The adversary can simply match the genotype data that is associated with the identity of the individual and the genotype data that is associated with the attribute. Such an attack requires only a small number of autosomal SNPs. Empirical data showed that a carefully chosen set of 45 SNPs is sufficient to provide matches with a [TYPE I ERROR] of $10^{-15}$ for most of the major populations across the globe[58]. Moreover, it is expected that random subsets of approximately 300 common SNPs would yield sufficient information to uniquely match any person[59].

With the low number of SNPs required for matching, individual level genotypes-phenotype records in genome-wide association studies (GWAS) are highly vulnerable to ADAD. In order to address this issue, several organizations, including the NIH, adopted a two tier access system for GWAS datasets: a restricted access area that stores individual level genotypes and phenotypes and a public access area for high level data summary statistics of allele frequencies of all cases and controls[60]. The premise of this distinction was that summary statistics enable secondary data usage for meta-GWAS analysis while it was thought that this type of data is immune to ADAD.

### ADAD: THE SUMMARY STATISTIC SCENARIO

A landmark work by Homer et al. reported the possibility of ADAD on GWAS datasets that only consists of the allele frequencies of the study participants[61]. The underlying concept of their approach is that, with the target genotypes in the case group, the average allele frequencies will be positively biased towards the target genotypes compared to the estimated MAF from the general population. Conversely, when the target is not part of the study, the average allele frequencies will be negatively biased compared to the target genotypes. A good illustration of this concept is considering an extremely rare variation in the subject's genome. Non-zero allele frequency of this variation in a small-scale study increases the likelihood that the target was part of the study, whereas zero allele frequency strongly reduces this likelihood. Homer et al. showed that by integrating the slight biases in the allele frequencies over a large number of SNPs it is also possible to conduct ADAD with the common variations that are analyzed in GWAS.

Subsequent studies extended the range of exploitations for summary statistics. One line of studies improved the test statistic in the original Homer et al. work and analyzed its mathematical properties[62-64]. Under the assumption of common SNPs in LINKAGE EQUILIBRIUM, their improved test statistic is mathematically guaranteed to yield maximal POWER for any SPECIFICITY level. Wang et al. went beyond allele frequencies and demonstrated that it is possible to exploit local LD structures for ADAD[65]. Their test statistic scores the co-appearance of two SNP alleles in the target genome with the bias of LD structure in a GWAS study versus the general population. The power of this approach stems from scavenging for the co-occurrence of two mildly uncommon alleles in different haplotype blocks that together create a rare event. They reported a power of 80% and specificity of 95% for ADAD on a GWAS with 200 samples that exploited the LD structure of 174 common SNPs in the FGFR2 locus. With the same number of SNPs, ADAD methods that use only allele frequencies yield an expected power of 24% for the same specificity level under the most optimal scenario. Im et al. developed a method to exploit the EFFECT SIZES of GWAS studies of quantitative traits to detect the presence of the target[66]. Different from ADAD with allele frequencies, the detection performance is better for participants with extreme phenotypes and worse for participants with average phenotypes. A powerful development of this approach is exploiting GWAS studies that utilize the same cohort for multiple phenotypes. The adversary repeats the identification process of the target with the effect sizes of each phenotype and integrates them to boost the identification performance. After determining the presence of the target in a quantitative trait study, the adversary can further exploit the GWAS data to predict the phenotypes with high accuracy[67]. This method works by simply correlating the DNA of the target with the effect sizes and takes advantage of the spurious associations when regressing a large number on markers with a single phenotype.

The theoretical performance of ADAD is a complex function of the size of the study and the general population[68,69]. On one hand, in any of the techniques above, studies with smaller numbers of participants generate more apparent biases in their summary statistics, which increases the power and specificity of the ADAD discrimination (**Figure 2A**). On the other hand, a target drawn randomly from the general population has a lower a-priori probability of having participated in a study with a smaller number of participants. This means that ADAD on smaller studies needs to work with higher specificity to achieve the same PRECISION of larger studies, reducing the power of the attack and the number of people at risk (**Figure 2B**). In any case, the performance and risk increase when the base population is smaller, such as the Amish or Hutterite populations, or when the meta-information enables stratification of the general population (**Figure 2C**).

The actual risk of ADAD on summary data has been the subject of debate. Following the original Homer et al. study, the NIH and other data custodians moved their GWAS summary statistic data from public databases to access controlled databases such as dbGAP[70]. A retrospective analysis found that significantly fewer GWAS studies publicly released their summary statistic data[71]. Most of the studies publish summary statistic data on 10-500 SNPs, which is compatible with one suggested guideline to manage risk[67]. Some warned that these policies are too harsh[72]. There are several practical complications that the adversary needs to overcome to launch a successful attack, such as access to the target's DNA data[73], access to a large reference database to assess the general population frequency data, and accurate matching between the ancestries of the target with those listed in the reference database[74]. Failure to address any of these prerequisites can severely impact the performance of the ADAD. In addition, for a range of GWAS studies, the associated attributes are not sensitive or private (e.g. height). Thus, even if ADAD occurs, the impact on the participant should be minimal. A recent NIH workshop proposed the release of summary statistics as the default policy and developing an exemption mechanism for studies with increased risk due to the sensitivity of the attribute or the vulnerability level of the summary data[75].

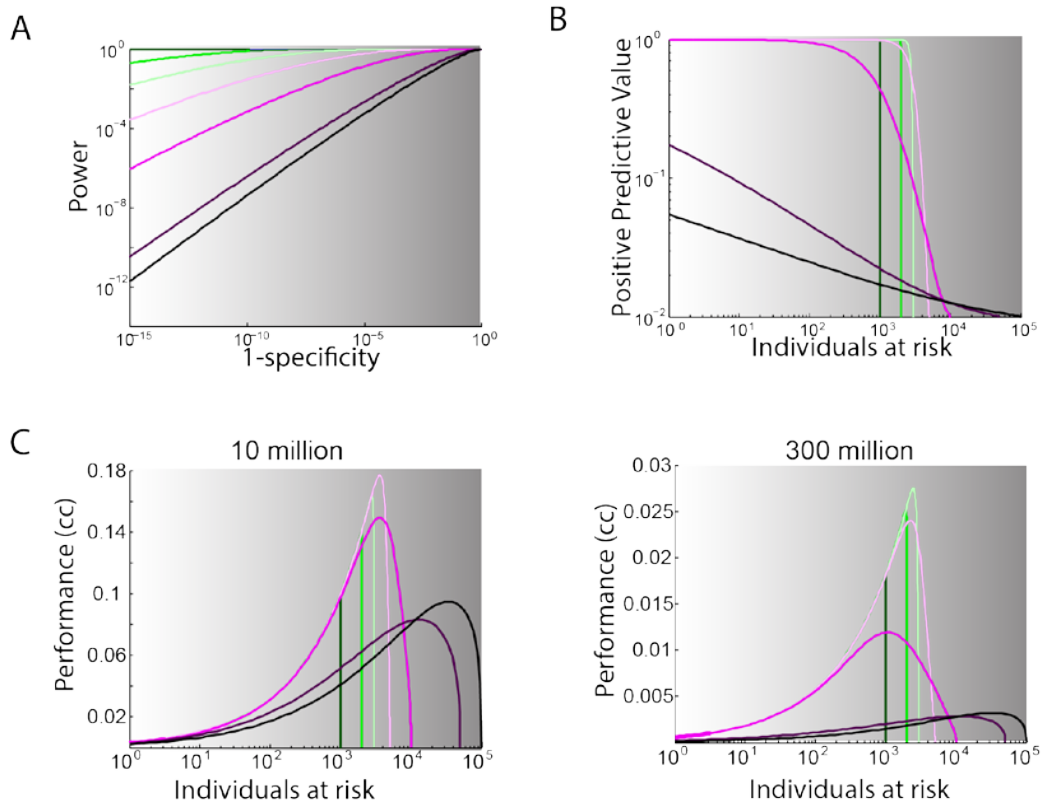

**Figure 2 | The performance of ADAD attacks using allele frequencies.** All conditions are with 50,000 common SNPs in linkage equilibirum. From dark green to dark purple, the lines denote studies with sizes: 103, 2×103, 3×103, 5×103, 104, 5×104, 105. The attribute is assumed to be positive in 1% of the base population in all conditions. a | The specificity and sensitivity of studies in different sizes with a fixed base population of 300 million people. b | The positive predictive value (the probabiltiy that a positive hit is truly positive) as a function of individuals at risk. Intermediate size studies risk the largest number of individuals for most of the positive predictive values. c | The ADAD performance (cc, correlation coeffcent between truth and prediction) as a function of individuals at risk for two base populations of 10 million and 300 milliom. For each base population, there is a study in a ceratin size that yields the best performance.

## ADAD: THE EXPRESSION DATA SCENARIO

Public databases such as GEO hold hundreds of thousands of gene expression profiles of individuals that are linked to a range of medical attributes. Schadt et al. proposed a potential route to exploit these profiles for ADAD[76]. The method starts with a training step that employs a standard EXPRESSION QUANTITATIVE TRAIT LOCI (eQTL) analysis with a reference dataset. The goal of this step is to identify several hundred strong eQTLs and to learn the distributions of the expression level for each genotype. Then, the algorithm scans the public expression profiles and calculates the probability distributions of the genotypes of the eQTLs. Last, the algorithm matches the target's genotype with the inferred allelic distributions of each expression profile and tests the hypothesis that the match is random. If the null hypothesis is rejected, the algorithm links the identity of the target to the medical attribute in the gene expression experiment. This ADAD technique has the potential for relatively high accuracy in ideal conditions. The method perfectly matched 580 individuals with their expression profiles when the training was conducted on a distinct dataset. Based on large-scale simulations, they further predicted that the method can reach a type I error of $1 \times 10^{-5}$ with a power of 85% when tested on an expression database using the entire US population.

There are several practical limitations to ADAD via expression data. While the training step and inference steps are capable of working with expression profiles from different tissues, the method reaches its maximal power when the training and inference utilize eQTL from the same tissue. Moreover, there is a significant loss of accuracy when the expression data in the training phase is collected using a different technology than the expression data in the inference phase. Another complication is that in order to fully execute the technique on a large database such as GEO, the adversary will need to manage and process large-scale expression data. Due to these practical barriers, the NIH did not issue any changes to their policies regarding sharing expression data from human subjects.

## COMPLETION ATTACKS

Completion of genetic information from partial data is a well-studied task in genetic studies, called genotype imputation[77]. This method takes advantage of the LINKAGE DISEQUILIBRIUM between markers and uses reference panels with complete genetic information to restore missing genotype values in the data of interest. The very same strategies enable the adversary to expose certain regions of interest where only partial access to the DNA data is available. One publicized case of a completion attack was the inference of Jim Watson's risk for Alzheimer's disease. Watson opted to publish his entire identified genome sequence except data from his ApoE gene, which is associated with Alzheimer's disease[78]. Nyholt et al. restored the ApoE status using imputation with markers that are 15Kb away from the masked site[79]. As a result of the study, a 2Mb segment around the ApoE gene was removed from Watson's published genome.

In some cases, completion techniques also enable the prediction of genomic sequences when there is no access to the DNA of the target. This technique is possible when the reference panel is combined with genealogical information[80]. The algorithm finds relatives of the target that donated their DNA to the reference panel and that reside on a unique path that includes the target, for example a pair of half-first cousins when the target is their grandfather. A shared DNA segment between the relatives indicates that the target had the same segment. By scanning more pairs of relatives that are connected through the target, it is possible to infer the two copies of autosomal loci and collect more genomic information on the target without any access to its DNA. Building on the deep genealogical records in Iceland, deCode Genetics was able to leverage their large reference panel to infer genetic variants of an additional 200,000 living individuals who never donated their DNA to the company. While this technique is mostly relevant to targets with a large number of decedents and can be executed in only a narrow range of scenarios, it emphasizes the complexities of genetic privacy. In May 2013, Iceland's Data Protection

Authority prohibited the use of this technique until consent can be obtained from the individuals who are not part of the original reference panel[81].

## MITIGATION TECHNIQUES

Most of the genetic privacy breaches presented above are quite sophisticated. They require a background in genetics and statistics and -- importantly -- a motivated adversary. One school of thought posits that these practical complexities almost eliminate the probability of an adverse event and therefore attenuate the risk to negligible levels for most studies[82,83]. According to this approach, an appropriate mitigation strategy is just removing very obvious identifiers from the datasets before publicly sharing the information. In the field of computer security, this risk management strategy is called security by obscurity. This approach is simple to implement and poses minimal burden on data dissemination. The opponents of security by obscurity posit that risk management schemes based on the probability of an adverse event are fragile and short lasting. According to their views, technologies only get better and what is technically challenging but possible today will be much easier in the future. Therefore, the probabilities of adverse events are non-computable and irrelevant[84]. Known in cryptography as Shannon's maxim[85], this school of thought assumes that the adversary exists and is equipped with the knowledge and means to execute the breach. Robust data protection, therefore, is achieved by explicit design of the data access protocol rather than by the actual chance of a breach[86]. This section surveys the main security by design schemes and their relevance to protecting genetic data.

### ACCESS CONTROL

One approach to mitigate the chance of a privacy breach is to place the sensitive data in a secure location and screen the legitimacy of the applicants and their research projects. Once approval is made, the applicants are allowed to download the data under the conditions that they will store it in a secure location and will not attempt to identify individuals. In addition, the applicants should be required to file periodic reports about the data usage and any adverse events. This approach is the cornerstone of the access-controlled dbGAP[60]. Based on periodic reports of the users, a retrospective analysis of dbGAP access control has identified 8 data management incidents in close to 750 studies, mostly non-adherence to the technical regulations, and no reports of breaching the privacy of participants[87]. Despite the absence of privacy breaches thus far, some have criticized the fact that access control creates an illusion of security[88]. Once the data is in the hand of the applicant, there is no real oversight of how it is being stored, the actual work, and what exactly is published. To address these limitations, an alternative approach is the trust-but-verify model, where the user cannot download the raw data but may execute certain types of queries that are recorded and monitored by the system[89]. Supporters of this model state that monitoring has the potential to deter malicious users from accessing the data and facilitates early detection. Another development based on this approach is enforcing the users and data custodians to have 'skin in the game'[90], by adding penalties beyond denying access to the resource in case of misuse. The main downside of access control is that any of the models listed above require constant management of the resource and create administrative burden to both data custodians and users.

## DATA ANONYMIZATION AND AGGREGATION

The premise of anonymity is the ability to be 'lost in the crowd'. One line of studies suggested restoring anonymity by restricting the granularity of the quasi-identifiers to the point that each record in the database is not unique. A popular heuristic is k-anonymity[91]. Using this approach, the quasi-identifiers are binned such that each subject's record is identical to that of at least k-1 records from other individuals in the dataset. To maximize the utility of the data for subsequent analysis, the binning process is adaptive. Certain records will have a lower resolution depending on the distribution of the other records and certain data categories that are too unique are suppressed entirely. There is a strong trade-off in the selection of the value of k; high values increase the size of the background crowd but at the same time reduce the utility of the data. As a rule of thumb, it was recommended to set k≥5 (92). More recent work showed that while k-anonymity protects against identity tracing techniques, it is vulnerable to attribute disclosure, especially when the adversary has a certain level of prior knowledge about the presence of the target in the database[93]. Subsequent studies developed more elaborative redaction techniques to address these issues[93,94]. These anonymization techniques have been mainly successful in safeguarding demographic identifiers in medical research. However, attempts to adopt these techniques to DNA research are yet to be practical[95]. The high dimensionality of DNA data dictates that most of the records will be unique and it is not clear how the data can be redacted without destroying its value for secondary analysis.

Differential privacy offers a distinct approach to restore anonymity by producing summary statistics after sophisticated data perturbation[96]. It aims to ensure that summary statistics of two datasets that differ by exactly one individual's record are extremely close to each other. This way, the adversary cannot be sure whether the target was part of the dataset or not and therefore cannot learn sensitive attributes. The challenge in differential privacy algorithms is to minimize the perturbation while satisfying the privacy property so that the summary statistic will still convey useful information on the population as a whole. Differential privacy has gained popularity in computer science and statistics as a very vibrant research area and the US Census Bureau uses this technique for their OnTheMap tool[97]. Early attempts have made progress towards protecting GWAS data using this approach[98,99]. Currently, the main limitation is that the amount of perturbation that needs to be added to the summary statistic grows linearly with the number of exposed SNPs, which quickly abolishes the ability to detect fine associations in meta-analysis. Whether or not there is a way to add much smaller amounts of noise in a way that still maintaining privacy for GWAS datasets remains an open question.

> **BOX 2 | Homomorphic encryption**
>
> Homomorphic encryption is a subfield in cryptography encryption with great potential for certain types of privacy preserving computation. It is best explained by the following analogy: Alice possesses raw gold and wants to create a necklace, but she is not equipped with the knowledge or tools. Bob is a skillful goldsmith but with an unclear reputation. Using homomorphic encryption, Alice sets up a securely locked glove box with the raw gold. Bob uses the gloves to construct the jewelry without unlocking the box. After that, Alice receives the glove box and opens the lock with her key. Similarly, genotypes can be thought of as the raw gold, Bob can be an interpretation service, and the necklace is disease risk status.
>
> Homomorphic encryption creates the glove box by adding additional mathematical properties besides the basic encryption and decryption operations in traditional cryptographic protocols. This property takes a regular function that operates on plaintext (genotypes), say $y(M_1,M_2)=M_1+M_2$, and finds a homolog function, $y'(X_1,X_2)$ that works on the ciphertext. Decrypting $y'(X_1,X_2)$ yields exactly the same answer as calculating the original function with the corresponding plaintext: $D(y'(X_1,X_2)) = M_1+M_2$, in our example. This way, Bob uses the homolog functions on the ciphertext and Alice decrypts and obtains the result.
>
> Until recently, cryptographic studies found protocols that could support functions with very basic cryptographic operations. One example is the Paillier Cryptosystem[110], which supports the addition of plaintext and multiplication by a constant. Such restricted implementations are called Partially Homomorphic Encryption. They operate relatively fast and despite their limitations, they might prove sufficient for a wide range of computations on genotypes due to the additive properties of many types of genetic risk predictions. A breakthrough in 2009 established a Fully Homomorphic Encryption[111] scheme that supports calculating a high order polynomial of the plaintext. This innovation is not yet efficient in terms of computational time but further developments can complete the arsenal of secure functions in genetic epidemiology.

## CRYPTOGRAPHIC SOLUTIONS

Modern cryptography brought new advancements for data dissemination beyond the traditional usage of encrypting sensitive files and distributing the key to authorized users. S*ecure multiparty computation* (SMC) allows two or more entities who each have some private data to execute a computation on these private inputs without revealing the input to each other or disclosing it to a third party. In one classical example of SMC, ALICE and BOB can determine who is richer without either one revealing their actual wealth to the other. Researchers have constructed SMC protocols in various domains—from voting[100] to location-based services[101].

In the area of genetic data, one line of work has developed SMC algorithms for genetic matching. Bruekers et al. presented a privacy-preserving algorithm to match STR profiles between two parties without exposing the actual genetic data[102]. Bohannon et al. suggested searchable genetic databases for forensic purposes that allow only going from genetic data to identity but not from identity to genetic data[103]. In their scheme, the records in the databases are encrypted with the individual's genotype as the key. To tolerate genotyping errors or missing data, they utilize a fuzzy encryption scheme that can use a key that only approximately matches the original one. This way, only access to the genotype information can reveal the identity but not the opposite. Along similar lines, Cristofaro et al. constructed cryptographic protocols for privacy-preserving paternity tests and genetic compatibility tests[104], albeit for molecular techniques that are no longer in use, such as RFLP. They also presented a smartphone-based implementation of these protocols[105]. The performance varies dramatically between tasks that examine only a few loci and those that depend on the whole genome. The former complete in under a second and the latter take days of computation and gigabytes of bandwidth, rendering them impractical at the current time.

In another direction, Kamm et al suggested a secure multi-center GWAS analysis[106]. In their protocol, each center deploys a secret sharing scheme on its own collection of subjects' phenotypes and genotypes that divides the data into small shares, each of which reveals nothing about the original values on its own. The shares are then sent to the other centers, which store them in dedicated servers. The servers have an interface that allows outsiders to initiate a GWAS study on phenotypes and genotypes of interest. Upon request, the servers coordinate to perform the association without reconstructing the original genotypes or phenotypes and only report in plain text the significant SNPs. A potential shortcoming of their approach is that, at least theoretically, the end product plain text is still vulnerable to ADAD on summary statistic data, rendering the solution far from complete.

Another line of cryptographic work looks at privacy-preserving outsourcing of computations on genetic information using homomorphic encryption[107] (**Box 2**). The concept of this approach is that, with advent of ubiquitous usage of genetic data, users (or physicians) will interact with a variety of genetic interpretation services (e.g. promethease.com) throughout their lives, which increases the chance of a genetic privacy breach. Under this cryptographic work, users send an encrypted version of their genome to the cloud. The interpretation service can access the cloud data but does not have the key and therefore cannot read the plain genotype values. Instead, the interpretation service executes the algebraic operations of its genetic risk prediction algorithm on the encrypted genotypes without inspecting the plaintext. After completing the algorithm, the user grabs the cyphertext results from the cloud. Due to the special mathematical properties of the underlying cryptosystem, the user simply decrypts the results to obtain his risk prediction. This way the user does not expose any of his genotypes or disease susceptibility to the service provider. The current scope of risk prediction models is still limited but this approach is quite amenable to future improvements.

# CONCLUSION

The invention of asymmetric cryptography in the 1970's led to a revolution in secure communication. Today, a wide variety of Internet transactions build upon these security measures in ways that are completely transparent to the average user. Data privacy still awaits a similar breakthrough. The status quo has greatly shifted in the last few years, with a torrent of studies showing that a motivated, technically-sophisticated adversary is capable of exploiting a wide range of genetic data for unintended purposes. With the constant innovation in genetics and the explosion of online information, we can expect that new privacy breaching techniques will be discovered in the next few years. Restoring the status quo with technical means will necessitate large strides in the theory and implementation of mitigation algorithms. Some of the approaches, particularly access control, have been quite useful. But so far, mitigation schemes are resource and time consuming for both the data custodian and users. Due to both technical and human factors[108], the privacy field has yet to come up with a set of methodologies of comparable impact to communication security.

Successful balancing of privacy demands and data sharing is not restricted to technical means[109]. Balanced informed consents outlining both benefits and risks are key ingredients for maintaining long-lasting credibility in genetic research. With the active engagements of a wide range of stakeholders from the broad genetics community and the general public, we as a society could develop social and ethical norms, legal frameworks, and educational programs to reduce the chance of misuse of genetic data despite the inability to theoretically prevent privacy breaches.

## GLOSSARY

SAFE HARBOR: A standard in the HIPAA Rule for de-identification of protected health information by removing 18 types of quasi-identifiers.
HAPLOTYPES: A set of alleles along the same chromosome.
CRYPTOGRAPHIC HASHING: A procedure that yields a fixed length output from any size of input in a way that is hard to determine the input from the output.
DICTIONARY ATTACKS: A brute force approach to reverse cryptographic hashing by scanning the relatively small input space.
TYPE I ERROR: The probability to obtain a positive answer from a negative item.
LINKAGE EQUILIBRIUM: Absence of correlation between the alleles in two loci.
POWER: The probability to obtain a positive answer for a positive item.
SPECIFICITY: The probability to obtain a negative answer for a negative item.
EFFECT SIZES: In quantitative traits, the contribution of a certain allele to the value of the trait.
EXPRESSION QUANTITATIVE TRAIT LOCI: Genetic variants associated with variability in gene expression.
LINKAGE DISEQUILIBRIUM: The correlation between alleles in two loci.
ALICE AND BOB: Common placeholders in cryptography to denote party A and party B.

## ACKNOWLEDGEMENTS

YE is an Andria and Paul Heafy Family Fellow and holds a Career Award at the Scientific Interface from the Burroughs Wellcome Fund. This study was also supported by a gift from Cathy and Jim Stone. The authors thank Dina Zielinski and Melissa Gymrek for useful comments and Shriram Sankararaman for his nice introduction between the authors.

## COMPETING INTERESTS STATEMENT

None.